\documentclass[5p,twocolumn]{elsarticle} 

\usepackage{graphicx}
\usepackage{nomencl} 

\usepackage[nodots,nocompress]{numcompress}

\usepackage{amssymb,amsmath}

\usepackage{lineno}

\journal{Int. J. Hydrogen Energy}
\makenomenclature 

\begin{document}

\begin{frontmatter}



\title{Models for Metal Hydride Particle Shape, Packing, and Heat Transfer}

\author{Kyle C. Smith}\ead{kyle.c.smith@gmail.com}
\author{Timothy S. Fisher\corref{cor1}}\ead{tsfisher@purdue.edu}
\address{Purdue University\\%
        School of Mechanical Engineering and Birck Nanotechnology Center\\%
        1205 W. State St.\\%
        West Lafafyette, IN 47907}
\cortext[cor1]{Corresponding author. Tel.: +1-765-494-5627; fax: +1-765-494-4731}

\begin{abstract}

A multiphysics modeling approach for heat conduction in metal hydride powders is presented, including particle shape distribution, size distribution, granular packing structure, and effective thermal conductivity.  A statistical geometric model is presented that replicates features of particle size and shape distributions observed experimentally that result from cyclic hydride decreptitation.  The quasi-static dense packing of a sample set of these particles is simulated via energy-based structural optimization methods. These particles jam (i.e., solidify) at a density (solid volume fraction) of $0.665 \pm 0.015$ -- higher than prior experimental estimates.  Effective thermal conductivity of the jammed system is simulated and found to follow the behavior predicted by granular effective medium theory.  Finally, a theory is presented that links the properties of bi-porous cohesive powders to the present systems based on recent experimental observations of jammed packings of fine powder.  This theory produces quantitative experimental agreement with metal hydride powders of various compositions.

\end{abstract}


\end{frontmatter}



\section{Introduction}

Metal hydrides offer high volumetric hydrogen storage density for on-board fuel cell vehicles \cite{SclNat2001}, and offer much potential for electrodes in electric vehicle batteries \cite{OvsSci1994}.  These metals undergo fragmentation induced by volumetric hydriding strain and embrittlement that results in the formation of irregular faceted particles \cite{CheJPDAP2007}.  Systematic reductions in average particle size have been observed as the number of hydriding cycles increases \cite{SakZPC2001,KojJAC2006}.  Also, hydrides can reach a state at which particles become mechanically stabilized, and the size distribution becomes invariant with further cycling \cite{SakZPC2001}.  This fragmentation process is essential to achieving fast hydriding kinetics, because the process exposes fresh chemically active surfaces \cite{FlkIJHE2010}, but the fragmented nature of metal hydride particles inhibits hydriding heat dissipation \cite{JinJHT2005}.  Expanded graphite \cite{KimIJHE2001,KleIJHE2004,RodIJHE2003} and metal \cite{RonJLCM1991,JosJLCM1984,BerJLCM1989} additives composited with metal hydrides can enhance heat dissipation during the hydriding process, but these chemically inactive materials are parasitic to hydrogen storage density and permeability.  In contrast, fragmented hydride packings exhibit high hydrogen storage density and permeability, but the dependence of effective thermal conductivity on the packed structure of hydride powders is not understood well.  In the present work, models are developed to address fundamental aspects of these particles, their packings, and their associated thermal properties to enable materials engineering.

\begin{figure}
\centering
\includegraphics[width=0.8\columnwidth]{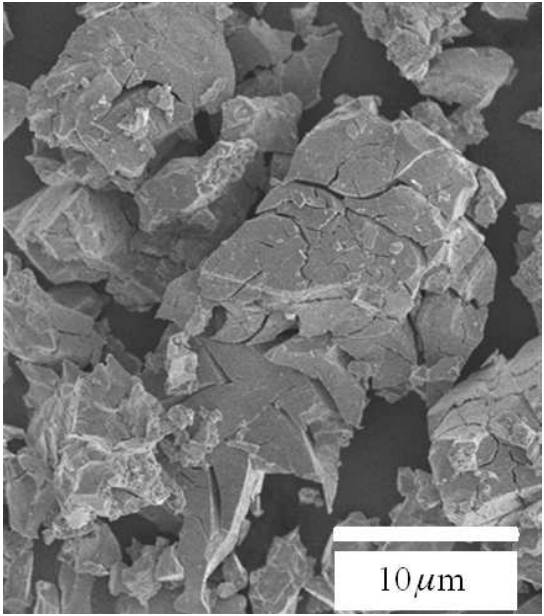}
\caption{Ti$_{1.1}$CrMn particles with crack fissures produced from cyclic hydriding and dehydring.  The particles produced after further cycling possess irregular faceted shapes.}
\label{fig:expfig}
\end{figure}

Hydrogenation causes metal hydride particles to expand, with approximately 2 to 5 \AA$^3$ per H atom \cite{Fuk2005}.  Assuming a density of 6.0 g/cm$^3$ for Ti$_{1.1}$CrMn and hydride composition Ti$_{1.1}$CrMnH$_2$ \cite{KojJAC2006}, volume expansion is expected to be approximately 9 to 23 \%.  Hydriding in metal particles requires diffusion of hydrogen through the lattice, with a hydride layer forming on the outer particle surface \cite{RudJAP1979}.  Thus, density mismatch at the metal-hydride interface induces large stresses on the material, inevitably leading to fracture.  Hahne and Kallweit \cite{HahIJHE1998} documented the dependence of particle size distribution on the number of hydriding cycles. They observed a five-fold reduction in average particles size after 30 cycles.  Particle morphology resulting from hydrogen-induced fracture can result in faceted particles having irregular shape.  This behavior has been observed for a wide variety of intermetallic metal hydrides for which particle size distributions have been measured (e.g., in Ref.~\cite{HahIJHE1998}), but no attempt has been made to predict the size distribution and shapes of irregular particles theoretically.

Particle shape resulting from hydrogen-induced fragmentation can result in faceted particles having irregular shapes (Fig.~\ref{fig:expfig}), in contrast to faceted particles produced from growth of single crystals.  The particular metal hydride used as an example here, Ti$_{1.1}$CrMn, is prepared by water-cooled arc melting \cite{KojJAC2006} that typically yields polycrystalline microstructures; disordered, polycrystalline alloy microstructures also have been shown to yield excellent hydriding characteristics \cite{OvsSci1994}.  Therefore, crystal defects and grain boundaries at which cracks initiate are expected to have random spatial and directional distribution.  Materials such as CeH$_{2.84}$ \cite{ManPRB2008} exhibit multiscale fractured structures as a result of phonon confinement energetics, but this phenomenon has not been observed for Ti$_{1.1}$CrMn.  

Strain-induced fragmentation is not restricted to systems undergoing hydriding and has been suggested as a universal phenomenon \cite{GroEPL1994}.  For example, thin layers of dried mud crack readily as a result of contraction and substrate friction \cite{GroEPL1994}.  The patterns formed by cracked mud exhibit remarkable resemblance to Si thin film anodes subjected to lithiation \cite{BeaESSL2001}.  More recently, investigations of lithiated Si nanopillars reveal that fracture planes have a highly anistropic directional distribution \cite{LeePNAS2012}.

Interparticle forces and macroscopic stresses result from this particle expansion and induce rearrangement and deformation of the powder.  Also, after initial cycling, the decrepitated powder is typically packed into a reactor, but no models for effective thermal conductivity have incorporated packing of these materials.  The influence of processing conditions on packed structure are important, because structures with enhanced heat conduction rates and active material density are highly desirable for compact, high-power fuel cell power systems.  By considering that packed hydride powder is a granular material, understanding of the phases formed by granular media may be applied to hydrides.  For instance, the jamming point represents the state at which a granular material develops rigidity and mechanical stability \cite{HerPRE2003}, and the jamming threshold density $\phi_J$ \nomenclature[phJ]{$\phi_J$}{jamming threshold density, -} is the density $\phi$ \nomenclature[ph]{$\phi$}{density, -} (i.e., solid volume fraction) at which this transition occurs.  Recently, we have introduced an energy-based formalism for repulsive contact that enables the rigorous simulation of jamming of arbitrarily-shaped faceted particles \cite{SmiPRE2010,SmiPRE2011,SmiPCCP2012}.  The structure of systems of tetrahedra prepared by such methods have shown excellent agreement with experiment (cf., \cite{SmiPRE2010,JaoPRL2010}).

Particle shape and size distributions resulting from fracture cannot be replicated by a simple single-particle unit cell model, such as those of Asakuma and co-workers \cite{AsaIJHE2004,UeokaIJHE2007} and Zehner, Bauer, and Schl\"{u}nder (see \cite{TsoCEP1987} for details).  Reduced conduction through the gas phase as a result of boundary scattering and interfacial impedance mismatches of thermal energy carriers depends on the confining pore geometry of the packed powder.  The gas-phase boundary scattering mechanism has often been interpreted as the Smoluchkowski effect in the packed bed literature \cite{TsoCEP1987}.  Also, metal hydride particles inherently exhibit small contact areas in the packed state, with resultant constriction effects on solid-state thermal energy carriers (i.e., electrons and phonons).  The degree of ballistic and diffusive conduction in the solid phase depends on the change in chemical composition of the metal hydride system due to hydrogenation and dehydrogenation.  The extent to which each of these mechanisms limits heat transfer in metal hydride powders, and other types of porous media in general, is not well understood.

In this work a comprehensive approach is introduced for modeling particle shape and size distribution in Sec.~\ref{shape}, quasi-static packing through the energy-based simulation of granular jamming in Sec.~\ref{packing}, and heat transfer through direct solution of the heat diffusion equation in the resulting heterogeneous microstructures in Sec.~\ref{heat_transfer}.  In Sec.~\ref{GSTT} a structure-transport theory inspired by recent experiments on fine powder is subsequently developed to relate non-cohesive jammed system properties to experimental packings with solid density below that of the non-cohesive jammed system.

\printnomenclature

\section{Particle Shape}\label{shape}

\subsection{Theory}

We attempt to replicate the particle size and shape distribution of decrepitated metal hydride powder with an idealized statistical geometric model.  The major underlying assumptions of the model follow: (1) infinitely extending planar surfaces are formed from instances of fracture and (2) planes of fracture have isotropic statistical orientation and position throughout the material.  This model can be described as a 3D Poisson plane field, though no such statistical field has yet been proposed in the literature to our knowledge.  The 3D Poisson plane field is closely related to the 2D Poisson line field that has been studied previously \cite{Sol1978}.  Both assumptions are well justified for polycrystalline metals, such as Ti$_{1.1}$CrMn.  Simulation of this geometric field of planes is accomplished through sequential sectioning of a unit cube centered at the origin with randomly oriented and positioned planes.  Random directions are determined through the generation of three random numbers $x_i$ \nomenclature[xi]{$x_i$}{uniformly distributed random variables with $i=1,2,3$, -} that are uniformly distributed on the interval [0,1] ($x_1$ for planar position; $x_2$ and $x_3$ for unit normal direction).

The random position $\boldsymbol{r}_{p}$ \nomenclature[rp]{$\boldsymbol{r}_{p}$}{random position vector, m} of a given plane can be expressed in terms of the parameter $x_1$, the plane unit normal vector $\hat{n}_p$ \nomenclature[np]{$\hat{n}_p$}{plane unit normal vector, -}, and the centroid $\boldsymbol{r}_{c}$ \nomenclature[rc]{$\boldsymbol{r}_{c}$}{cube centroid position vector, m} of the cubic-shaped domain undergoing subdivision:

\begin{equation}
  \boldsymbol{r}_{p} = x_1 \hat{n}_p \sqrt{3}/2 + \boldsymbol{r}_{c}.
\end{equation}

\noindent The normal vector is expressed in terms of continuously distributed parameters, $x_1$, $x_2$, and $x_3$, assuming isotropic planar orientation, but the parameters could easily incorporate anisotropy by restricting them to a discrete set that reflects rotational symmetry of the underlying atomic structure (e.g., cubic or icosahedral).  The present model may be applicable to the pulverization of Si anodes by incorporating anisotropic distributions consistent with those presented in Ref.~\cite{LeePNAS2012}.  The probability $p(d\omega)$ \nomenclature[p]{$p$}{probability, -} for fracture along planes having normal vectors subtending a differential solid angle $d\omega$ \nomenclature[om]{$\omega$}{solid angle, steradians} about the direction $\hat{n}_p$ can be expressed in terms of azimuthal and zenith angles in the spherical coordinate system, $\theta$ \nomenclature[th]{$\theta$}{azimuthal angle, radians} and $\psi$ \nomenclature[ps]{$\psi$}{zenith angle, radians}, as well as the probability density $c$ \nomenclature[c]{$c$}{probability density, 1/steradians}:

\begin{equation}
  p(d\omega) = c d\omega = c d\theta d\psi \sin{\psi}.
\end{equation}

\noindent In the present study isotropic planar orientation is considered, and therefore $c$ is considered a constant independent of $\hat{n}_p$.  Because uniformly distributed independent random variables ($x_1$, $x_2$, and $x_3$) are used to describe direction, the differential solid angle $d\omega$ must be recast with $\delta = \cos{\psi}$ \nomenclature[de]{$\delta$}{zenith angle transformation variable, -}; this substitution guarantees that $p(d\omega)$ is proportional to $d\theta$ and $d\delta$, each having uniform distributions.  By defining $\delta$ and $\theta$ in terms of the parameters $x_2$ and $x_3$, the spherical coordinates specifying planar orientation can be expressed as:

\begin{equation}
  (\theta,\psi) = (2 \pi x_2,\cos^{-1}(2x_3-1)).
\end{equation}

\noindent With these spherical coordinates, the planar normal vector can be expressed:

\begin{equation}
  \hat{n}_p = \cos{\theta}\sin{\psi}\hat{i}+\sin{\theta}\sin{\psi}\hat{j}+\cos{\psi}\hat{k}.
\end{equation}

\begin{figure}
\centering
\includegraphics[width=\columnwidth]{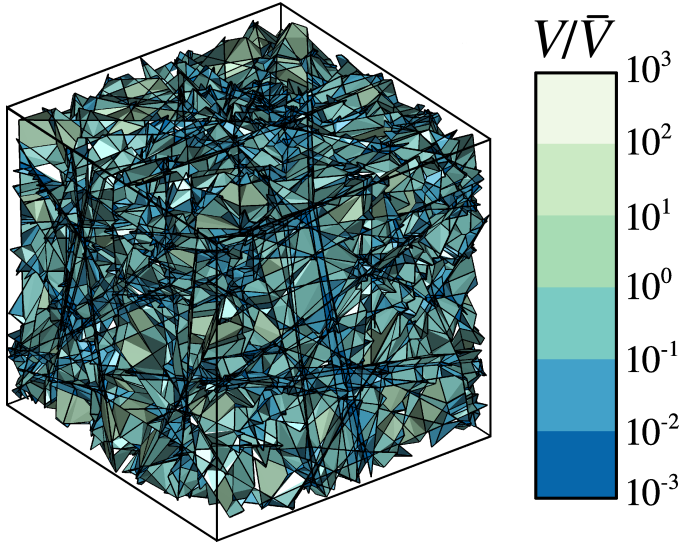}
\caption{Particles generated by intersections of planes in the 3D Poisson field.  Particles are colored according to their volume $V$ relative to the volume-weighted average particle volume $\bar{V}$ for the ensemble.  Particles intersecting the domain boundary (black edges) are excluded to neglect edge effects.}
\label{fig:fullvc}
\end{figure}

\nomenclature[ijk]{$\hat{i}$, $\hat{j}$, $\hat{k}$}{Cartesian unit vectors, -}

This model was implemented numerically, and can be interactively accessed through nanohub.org \cite{rndmfrc}.  Additional functionality exists in this tool that is outside the scope of this work; some of the extra tool functionality is described in our original work on this topic \cite{SmiMRS2009}.  The resulting ensemble of particles formed by the intersection of 200 planes in the Poisson field is depicted in Fig.~\ref{fig:fullvc}.  The volume of individual particles in the ensemble spans nearly six orders of magnitude.  Qualitatively, the field reflects many of the features of decrepitated metal hydride particles (cf., Fig.~\ref{fig:expfig}), including faceted particle shapes.  Comparison of the resulting particle size distributions with experimental data in the subsequent section also confirms excellent quantitative agreement of this simple statistical geometric model.

\subsection{Experiment}

\begin{table*} 
\centering
\caption{Details of metal hydrides considered.}
\begin{tabular}{cccc}
\hline
\hline
  Metal hydride                      & History           & $D_{50}$ ($\mu$m) & Reference \\
\hline
  MmNi$_{3.5}$Co$_{0.7}$Al$_{0.8}$   & 650 cycles        & $12$              & \cite{SakZPC2001} \\
  LaNi$_{4.7}$Al$_{0.3}$             & 30 cycles         & $7.2$             & \cite{HahIJHE1998} \\
  HWT                                & 11 cycles         & $6.3$             & \cite{HahIJHE1998} \\
  Ti$_{1.1}$CrMn                     & cycling/oxidation & $3.6$             & present \\ 
\hline
\hline
\end{tabular}
\label{table:cyclehistory}
\end{table*}

\nomenclature[D]{$D$}{Volume effective sphere diameter,  m}

The particle size distribution of a high pressure metal hydride, Ti$_{1.1}$CrMn, was measured with laser diffraction by Malvern Instruments \cite{Malvern}.  Prior to the measurements the material had undergone a process of cycled hydriding and dehydriding to chemically activate the material, as described in \cite{FlkIJHE2010}.  The pyrophoric tendency of Ti$_{1.1}$CrMn required controlled oxidation to passivate the exterior surfaces of metal hydride particles for exposure to air during particle size distribution measurements.  

\begin{figure}
\centering
\includegraphics[width=\columnwidth]{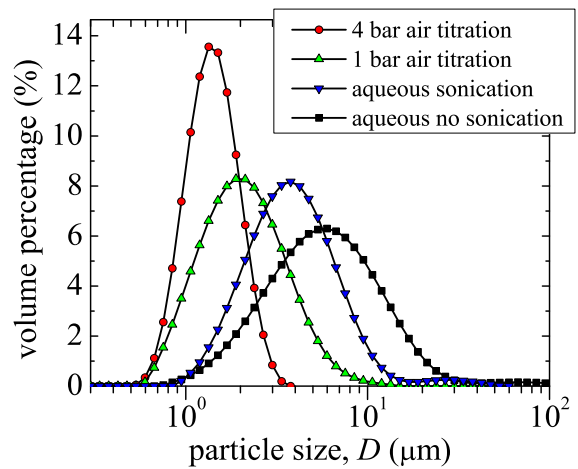}
\caption{Particle size distributions of passivated Ti$_{1.1}$CrMn by laser diffraction \cite{Malvern}.  Air titration (dry) and aqueous (wet) methods were used to disperse particles.}
\label{fig:ticrmnpsd}
\end{figure}

Particles were dispersed to obtain accurate measurements of particle size distributions by laser diffraction.  Both wet and dry methods were utilized and are outlined in ISO13320-1 \cite{Malvern}.  In the wet method particles were dispersed in water with and without ultrasonic excitation for subsequent analysis; this method is primarily intended to assess particle size in the absence of agglomeration \cite{Malvern}.  For sonicated samples, the particle size distribution was measured as a function of sonicating time to determine when agglomerates were completely fragmented.  Three minutes of ultra-sonication were required in order to achieve adequate dispersion of the passivated particles.  In the dry method particles were accelerated in air by a pressure differential induced across a venturi in which shear stresses act to reduce agglomeration, but this process may have induced attrition of particles \cite{Malvern}.  The resulting particle size distributions are displayed in Fig.~\ref{fig:ticrmnpsd}.  

The results indicate that the wet method yields larger particle sizes when sonication is not used; the median diameter was 3.8 and 5.7 $\mu$m with and without sonication, respectively.  Despite the reduction in measured particle size upon sonication, agglomerates still appear to be present, as reflected by a small secondary peak near 20 $\mu$m.  In contrast, the dry method results in much smaller median particle diameters of 1.4 and 2.0 $\mu$m for 4 and 1 bar applied pressure, respectively.  Also, sparks were observed during acceleration of particles at the highest pressure \cite{Malvern}, indicating that particles had undergone attrition, resulting in fresh pyrophoric surfaces.  These results support the notion that metal hydride particles possess an abundance of internal cracks even after the cyclic fragmentation process stabilizes.  Such microstructure within particles must be accounted for in the modeling of hydrogen diffusion kinetics and heat flow in metal hydride powder.

\begin{figure}
\centering
\includegraphics[width=\columnwidth]{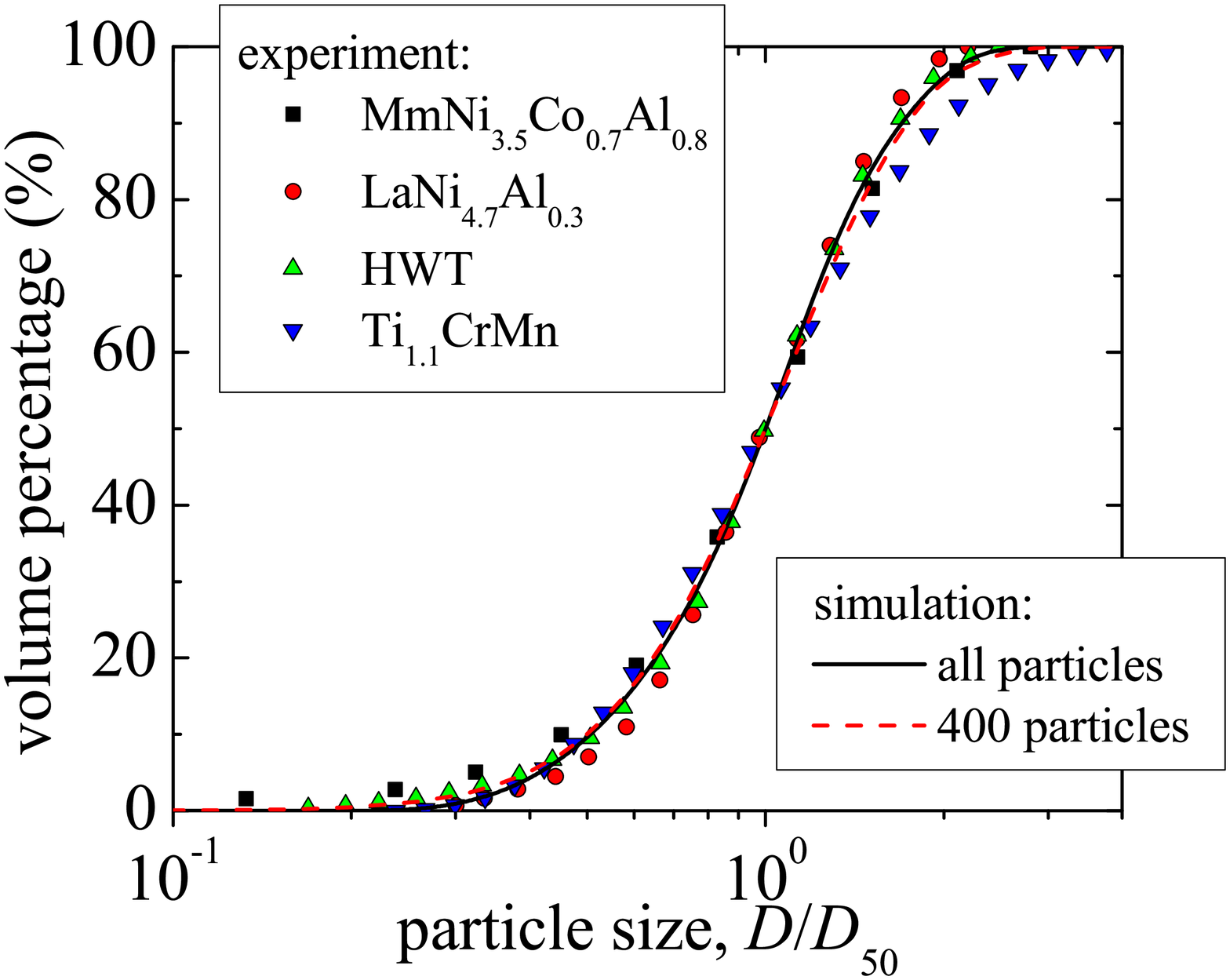}
\caption{Modeled and experimental cumulative size distributions for cycled metal hydride powder from the present work and Refs.~\cite{SakZPC2001,HahIJHE1998}.  Volume effective sphere diameter $D$ is plotted for theoretical size distributions.  $D$ is normalized by its median value $D_{50}$.  Kernel density estimates of the simulated distributions are shown.}
\label{fig:allcsd}
\end{figure}

The cumulative size distribution of aqueous sonicated Ti$_{1.1}$CrMn is displayed in Fig.~\ref{fig:allcsd} in addition to those of several other chemically dissimilar metal hydrides from the literature.  For the aqueous sonicated sample of Ti$_{1.1}$CrMn the weak secondary peak of the size distribution resulting from agglomeration was excluded.  Despite composition contrast, each metal hydride exhibits similar particle size distribution.  The most substantial deviation from similarity is for particles at the large end of the size range; this incongruence between measurements and model is likely due to the presence of agglomerates.  In addition, the various metal hydride samples had undergone various degrees of cycling, as shown in Table~\ref{table:cyclehistory}.  The lack of a physically consistent trend of particle size with the degree of cycling among the various hydrides in Table~\ref{table:cyclehistory} suggests that the median size of fully pulverized particles is primarily dependent on chemical composition and mechanical properties.  

Metal hydride particles generated from a Poisson plane field formed by the intersection of 200 randomly oriented and positioned planes are compared to experimental size distributions of the four different metal hydrides in Fig.~\ref{fig:allcsd}; 200 random planes were sufficient to converge the size distribution shape in Fig.~\ref{fig:allcsd}.  The size distribution of the Poisson plane field lies in the middle of the spread of experimental size distributions, indicating excellent agreement with the data.  Such a finding is remarkable considering that a simple model lacking adjustable parameters yields a particle size distribution that closely matches experimental observations.  This result suggests that the random nature of this process results in particle shapes that are relatively independent of material composition.

\section{Packing Theory and Simulation}\label{packing}

400 particles from the Poisson plane field are randomly sampled to study athermal (i.e., low particle kinetic energy) jamming of frictionless metal hydride-like particles; the methods employed to find the jammed structure of these particles are briefly described here.  The cumulative size distribution for this subset of particles closely matches that of the full Poisson plane field (Fig.~\ref{fig:allcsd}).  To simulate jamming, these particles are arranged in a dilute setting with random particle position and orientation in a periodic cubic supercell.  Sequentially, isotropic affine compressive strain is applied to the assembly.  Structural relaxation of particles is simulated at each density $\phi$ (\textit{i.e.}, solid volume fraction) via the minimization of elastic energy between contacting particles.  A contact mechanics model is employed as in \cite{SmiPRE2010,SmiPRE2011,SmiPCCP2012} for packings of Platonic solids and LiFePO$_4$ particles, where the contact energy $E_{\alpha\beta}$ between particles $\alpha$ and $\beta$ is:

\nomenclature[EAB]{$E_{\alpha\beta}$}{energy between particles $\alpha$ and $\beta$, J}

\begin{equation}
  E_{\alpha\beta} = \frac{YV_{\alpha \cap \beta}^{2}}{2(V_{\alpha}+V_{\beta})},
\end{equation}

\noindent and $Y$ is Young's modulus \nomenclature[Y]{$Y$}{Young's modulus, Pa}, $V_{\alpha \cap \beta}$ \nomenclature[VAB]{$V_{\alpha \cap \beta}$}{intersection volume between particles $\alpha$ and $\beta$, m$^3$} is the volume of intersection of the two undeformed polyhedral particles in contact, and $V_{\alpha}$ \nomenclature[VA]{$V_{\alpha}$}{volume of particle $\alpha$, m$^3$} is the volume of particle $\alpha$.  The present mechanics model permits computation of energy, forces, and moments for contact between two arbitrarily-shaped dissimilar particles; the generality of this model is especially important for the present study because of the random shape of particles in the Poisson plane field.  Detailed expressions for forces and moments and the numerical methods employed for structural optimization are described in Ref.~\cite{SmiPRE2010}.

Following the procedure outlined above, an initially dilute configuration of particles at $\phi = 0.005$ was consolidated sequentially until a stable state was reached at $\phi = 0.856$ with energy per particle of $1.5 \times 10^{-4}Y\bar{V}$, where $\bar{V}$ \nomenclature[Vb]{$\bar{V}$}{volume-weighted average particle volume, m$^3$} is the volume-weighted average particle volume.  The system was subsequently expanded to approach the jamming density, but the ill-conditioned nature of this system prohibited optimization to stable states near the jamming point.  Subsequently, the system was expanded to a density of $\phi=0.643$ and optimized to nearly zero energy, confirming that the jamming threshold density $\phi_J$ exceeds 0.643.  Further consolidation with fine strain stepping resulted in stagnant optimization progress near $\phi=0.65$, while estimates of the jamming threshold density based on limited contact depth data yield a jamming density of $0.68$ (see Ref.~\cite{SmiPRE2010} for this approach).  Based on these findings, we estimate the jamming density for this system to be $0.665\pm0.015$.

\begin{table*} 
\centering
\caption{Metal $\phi_M$ and metal hydride $\phi_{MH}$ \nomenclature[phM]{$\phi_{M(H)}$}{metal (hydride) packed density, -} packed densities (i.e., solid volume fraction).  Hydriding expansion was estimated at 2 to 5 \AA$^3$ for each hydrogen atom present in the metal (see Ref.~\cite{Fuk2005}).  Uncertainty bounds of $\phi_{MH}$ are expressed in terms of the uncertainty in volumetric expansion during hydriding.}
\begin{tabular}{ccccc}
\hline
\hline
  Metal hydride           & Expansion \%   & $\phi_M$ &   $\phi_{MH}$      & Reference \\
\hline
  Ti$_{1.1}$CrMn          & $16 \pm 7$     & 0.3      &   $0.35 \pm 0.02$    & \cite{FlkIJHE2010} \\
  LaNi$_{4.7}$Al$_{0.3}$  & $23 \pm 10$    & 0.469    &   $0.577 \pm 0.047$  & \cite{HahIJHE1998} \\
  HWT                     & $24 \pm 10$    & 0.555    &   $0.688 \pm 0.056$  & \cite{HahIJHE1998} \\
\hline
\hline
\end{tabular}
\label{table:expdensities}
\end{table*}

As displayed in Table~\ref{table:expdensities}, the jamming threshold density of this frictionless, non-cohesive system is higher than packed densities of experimentally prepared metal powder.  The act of hydriding expands particles and, therefore, increases the density of powder confined by a container of fixed size.  Estimates of the resultant hydride powder densities are included in Table~\ref{table:expdensities}.  The results suggest that the HWT hydride was packed near jamming, while the other samples were very loosely packed relative to the simulated jamming point.  Variation of these packed densities further suggests that cohesive van der Waals forces impede densification during packing (see \cite{YanPRE2000} and references therein for example).  Also, wet consolidation or mechanical vibration may be necessary to pack hydride near jamming and accordingly achieve jammed volumetric hydrogen storage densities.  We expect that the present jammed systems without friction and cohesion represent the densest \textit{random} packing of such particles \cite{PPFtesnote}.

\begin{figure}
\centering
\includegraphics[width=\columnwidth]{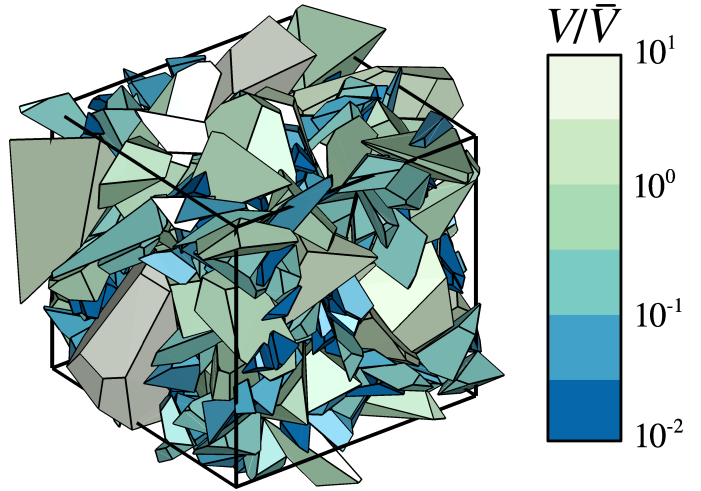}
\caption{Jammed packing of 400 particles randomly sampled from the Poisson plane field at $\phi=0.633$.  Particle color indicates volume $V$ relative to the volume-weighted average particle volume $\bar{V}$.  The cube-shaped periodic supercell is indicated by black edges.}
\label{fig:vcconfig}
\end{figure}

The jammed microstructure obtained through the aforementioned simulation is depicted in Fig.~\ref{fig:vcconfig} and reveals that small particles are well distributed and readily contact other small particles.  Conversely, large particles may be isolated from other large particles and at most will only contact two other large particles.  Visual inspection reveals that the full distribution of particles is engaged in the contact network in a non-trivial fashion.  In addition to the system of 400 particles, a system of 100 particles was also studied.  This ensemble was too small to reflect the structure in the large system limit, because unrealistic structural correlation was induced as a result of the similar size of the supercell relative to the largest particle in the system.  Though only a single realization of each system size is studied here, convergence of structural and transport properties among multiple realizations and sizes is expected as long as the jammed cell size exceeds two length scales of the maximal particle.

\section{Heat Transfer}\label{heat_transfer}


Transfer of heat by conduction occurs in materials via the flow of energy carriers, e.g., phonons, electrons, and molecules in semiconducting, metallic, and gas phases, respectively.  For transport in large systems with long time scales, diffusive processes are characterized by frequent isotropic scattering of carriers via intrinsic mechanisms (e.g., through carrier-carrier interactions).  Accordingly `large' systems with `long' times are rational only when compared to length and time scales of diffusive processes.  In the present study of steady-state heat transfer, only the diffusive length scale, the mean free path between scattering events $\lambda_{mfp}$, \nomenclature[lamfp]{$\lambda_{mfp}$}{mean free path, m} must be considered.  When the length scale through which heat is transported is smaller than $\lambda_{mfp}$, carriers are constricted only by boundaries, and the transport regime is referred to as ballistic.  Multiscale features present in granular materials can result in ballistic and diffusive behavior throughout the heterogeneous medium.  The reduction of heat flow through the gas phase occurs as a result of interfacial interactions of thermal energy carriers with the confining pore geometry of the granular medium.

\subsection{Theory}

The so-called Kapitza resistance and Smoluchkowski effects are known to occur in fluid-solid systems.  The Kaptiza resistance refers to a temperature jump resulting from the finite density of fluid molecules transferring heat normal to the interface \cite{SwaRMP1988}.  In contrast, the Smoluchowski effect involves the reduction of effective thermal conductivity of a gas as a result of molecular accommodation on the interface \cite{KenKTG1964}; this process can also be understood according to thin film theory as a boundary scattering effect \cite{SonAP1952}.  It is not clear, as yet, whether a simple Kaptiza resistance is sufficient to replicate the boundary scattering of gas molecules in porous media, i.e., the Smoluchowski effect.  We therefore consider both effects to understand the connections and limitations of these phenomena in realistic porous media.  Kaptiza resistance results from interactions between solid-gas interfaces oriented normal to the net heat flux $\vec{q}''$, \nomenclature[q]{$\vec{q}''$}{heat flux vector, W/m$^{2}$} whereas interfaces aligned to the transport direction scatter molecules, as a result of the Smoluchowski effect.

To assess whether the Kaptiza resistance effect or the Smoluchowski effect dominates in the transport process, we introduce a Biot number $Bi$ \nomenclature[Bi]{$Bi$}{Biot number, -}, similar to \cite{IncFHMT2007}, as the ratio of bulk thermal resistance in the gas to interfacial resistance at the solid-gas interface:

\begin{equation} 
  Bi = \frac{\delta_p}{R''\kappa_g},
\label{eq:Bi}
\end{equation} 

\noindent where $R''$ \nomenclature[RZ]{$R''$}{Kaptiza resistance, m$^2$-K/W} is the unit area Kaptiza resistance, $\kappa_g$ is the reduced gas thermal conductivity resulting from the Smoluchowski effect, and $\delta_p$ \nomenclature[dep]{$\delta_p$}{characteristic pore size, m} is the characteristic size of the pore.  We express this quantity as a function of Knudsen number $Kn$ \nomenclature[KZn]{$Kn$}{Knudsen number, -} defined as $\lambda_{mfp}/\delta_p$ and consider diffuse interfacial models for both effects.  For the Kaptiza resistance resulting from diffuse scattering of H$_2$ molecules at solid interfaces, we employ the approximation of the diffuse mismatch model \cite{SwaRMP1988}, where $R''$ is:

\begin{equation} 
  R''= \frac{4}{Cv}.
\label{eq:Gpp}
\end{equation} 

\noindent $C$ \nomenclature[CZ]{$C$}{volume-based specific heat, J/m$^3$} and $v$ \nomenclature[v]{$v$}{mean molecular velocity, m/s} are the specific heat (volume-based) and mean velocity of molecules, respectively.  For the reduced gas thermal conductivity $\kappa_g$ \nomenclature[kag]{$\kappa_g$}{gas conductivity, W/m-K} we consider an analogous form of the result presented by Sondheimer for electron transport in thin films with diffuse interfaces \cite{SonAP1952}:

\begin{equation} 
  \kappa_g= \frac{\kappa_0}{(1+3Kn/8)},
\label{eq:kapg}
\end{equation} 

\noindent where $Kn$ is Knudsen number and $\kappa_0$ \nomenclature[ka0]{$\kappa_0$}{bulk gas conductivity, W/m-K} is the bulk gas thermal conductivity.  Eq.~\ref{eq:kapg} is valid for $Kn<<1$.  Here the bulk gas conductivity $\kappa_0$ is expressed via kinetic theory as $\kappa_0=Cv\lambda_{mfp}/3$ \cite{KenKTG1964}.  Substitution of Eqs.~\ref{eq:Gpp} and ~\ref{eq:kapg} into Eq.~\ref{eq:Bi} and subsequent simplification yields:

\begin{equation} 
  Bi = \frac{3(1+3Kn/8)}{4Kn}.
\end{equation} 

\noindent At low pressure (i.e., $Kn \sim 1$) interfacial and volumetric resistances are similar because $Bi$ is near unity, while in the opposite limit of high pressure ($Kn \rightarrow 0$) $Bi$ diverges to infinity.  Thus, thermal resistance in the gas phase is large relative to that of the interface over much of the practical pressure range.  The Smoluchowski effect is therefore primarily required to represent heat transfer in packed metal hydride beds; only at very low pressures are Kaptiza resistances comparable.

The above analysis is useful for understanding the relative importance of each boundary interaction mechanism.  A directionally resolved solution of molecular convection and scattering within the gaseous region of the heterogeneous metal hydride powder is required to rigorously simulate the Smoluchowski effect in that medium.  Methods for such analyses (e.g., Boltzmann transport equation or Monte Carlo) have yet to be performed in the metal hydride literature, mainly because of the high computational expense and unknown microscopic pore structure.  In the present work, Fourier heat conduction in the gas and solid regions of the heterogeneous powder is simulated.  Later, a reduced gas conductivity model is incorporated to replicate the Smoluchowski effect in a manner similar to prior work (see \cite{DaWJLCM1990,AsaIJHE2004,TsoCEP1987,PonsJLCM1991,RodIJHE2003}).

\subsection{Simulation}\label{htsim}

\subsubsection{Methods}

We consider a heterogeneous granular medium composed of gas and solid with thermal conductivities $\kappa_g$ and $\kappa_s$, \nomenclature[kas]{$\kappa_s$}{solid conductivity, W/m-K} respectively.  The steady heat diffusion equation governs the microscopic temperature field $T$ \nomenclature[T]{$T$}{temperature field, K} in the medium:

\begin{equation}
 \nabla \cdot (\kappa \nabla T) = 0,
 \label{eq:heateq}
\end{equation} 

\noindent where $\kappa$ \nomenclature[ka]{$\kappa$}{local conductivity, W/m-K} is local conductivity.  The effective conductivity tensor $\boldsymbol{\kappa}$ \nomenclature[Kb]{$\boldsymbol{\kappa}$}{effective conductivity tensor, W/m-K} for the macroscopic medium relates the homogenized temperature gradient $(\nabla T)_h$ to the average heat flux vector $\langle \vec{q}'' \rangle$:

\begin{equation}
 \langle \vec{q}'' \rangle = -\boldsymbol{\kappa}(\nabla T)_h.
 \label{eq:kappatensor}
\end{equation}

\noindent Periodic temperature fall boundary conditions are applied to the granular medium to probe $\boldsymbol{\kappa}$:

\begin{equation}
  T(\boldsymbol{r})=T(\boldsymbol{r}+\boldsymbol{r}_0)+ (\nabla T)_h \cdot \boldsymbol{r}_0.
\end{equation}

\noindent $\boldsymbol{r}$ \nomenclature[r]{$\boldsymbol{r}$}{position in the heterogeneous medium, m} is position in the heterogeneous medium, $\boldsymbol{r}_0$ \nomenclature[r0]{$\boldsymbol{r}_0$}{integer multiple of lattice vectors, m} is a linear combination of integer multiples of principal lattice vectors defining the periodic supercell, and $(\nabla T)_h$ is the homogenized temperature gradient.  Three independent boundary values problems are solved with unit temperature drops in each of the principal Cartesian directions (i.e., $(\nabla T)_h=\hat{i}$, $\hat{j}$, and $\hat{k}$) to determine the components of $\boldsymbol{\kappa}$ based on its definition in Eq.~\ref{eq:kappatensor}.  Mean effective conductivity $\bar{\kappa}$ \nomenclature[kab]{$\bar{\kappa}$}{mean effective conductivity, W/m-K} is computed as the mean of the principal conductivities of $\boldsymbol{\kappa}$.

\begin{figure}
\centering
\includegraphics[width=\columnwidth]{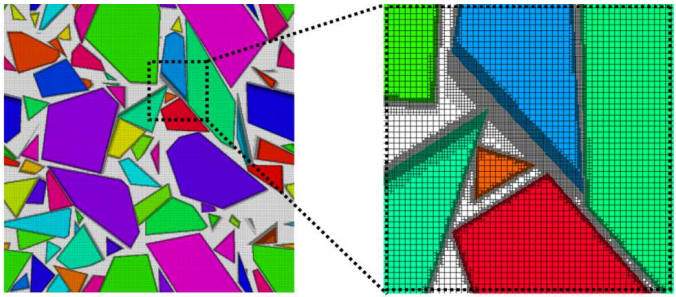}
\caption{The two dimensional projection of a robust octree-based mesh for the simulation of transport within the granular microstructure.  This system is composed of 400 irregular, polydisperse jammed particles, each of which exhibit a distinct color.  The mesh shown exhibits three levels of refinement near particle boundaries that is necessary to resolve fine gaps between particles.}
\label{fig:octreevc}
\end{figure}

To solve Eq.~\ref{eq:heateq} numerically on the heterogeneous domain, an approximate, robust meshing scheme is employed, wherein reconstructed interfaces are formed by that of a non-conformal multilevel Cartesian mesh.  To utilize computer memory efficiently while achieving adequate solid-gas interface resolution, octree-based refinement is employed on Cartesian cells intersecting solid-gas interfaces.  An example mesh employed to discretize the system of 400 three-dimensional irregular, polydisperse jammed particles is depicted in Fig.~\ref{fig:octreevc}.  

Refinement of the octree mesh was performed to verify adequacy of interfacial resolution.  At densities approaching jamming ($\phi \rightarrow \phi_{J}^{-}$), a high degree of refinement is required to resolve gas-phase gaps between solid particles approaching face-face contact.  For meshes having fine levels larger than these gaps, artificial continuity of solid phase can occur; to prevent this effect, cells exhibiting such artificial continuity are treated as gas phase.  In the limit of infinite refinement, the number of such cells vanishes, and consequently the present scheme is topologically consistent with jammed structures.  

In practice a coarse level mesh was used with cell size $\Delta x = 0.004L$ \nomenclature[Dx]{$\Delta x$}{finite volume cell size, m} for  \nomenclature[L]{$L$}{side length of primary supercell, m} a system of 400 particles at $\phi=0.633$, where $L$ is the side length of the periodic supercell.  Only 1\% deviation in $\bar{\kappa}$ was observed among meshes having two and three levels of refinement with $\kappa_s/\kappa_g = 10^3$, and therefore a two-level mesh was used to obtain the present results.  The two-level mesh contained $3.5\times10^7$ fine and $1.2\times10^7$ coarse cells.  The aggregation-based algebraic multigrid method \cite{NotayETN2010,NotayTech2010,NotayTech2011} was employed to iteratively solve the corresponding discrete set of equations with the implementation of Notay and co-workers.

\subsubsection{Results}

\begin{figure}
\centering
\includegraphics[width=\columnwidth]{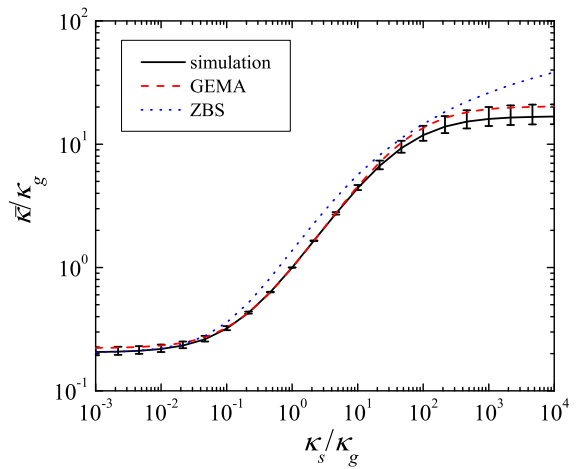}
\caption{Mean effective conductivity as a function of solid-fluid conductivity ratio $\kappa_s/\kappa_g$ for the jammed system at $\phi = 0.633$.  Error bars represent bounds of the principal values of the effective conductivity tensor $\boldsymbol{\kappa}$.  Theoretical data presented are based on the granular effective medium approximation (GEMA) in \cite{YonJAP1983,SenGeo1981} and the Zehner-Bauer-Schl\"{u}nder (ZBS) model (see \cite{TsoCEP1987}).}
\label{fig:kks}
\end{figure}

The mean effective conductivity $\bar{\kappa}$ of the jammed system at a density just below the jamming point ($\phi=0.633$) was computed for a range of solid-gas conductivity ratios $\kappa_s/\kappa_g$, as displayed in Fig.~\ref{fig:kks}.  $\bar{\kappa}/\kappa_g$ approaches an asymptote for high solid conductivities.  This behavior implies that the solid phase does not form a high-conductivity path through which heat preferentially flows.  The structure of the granular medium induced by the jamming of these highly irregular, faceted particles is responsible for this behavior.  Heat is forced to flow through the gas phase in order to reach highly conductive solid islands.  Such poor solid continuity is consistent with observations in the literature of very low effective conductivity of evacuated metal hydride powder (see \cite{HahIJHE1998}).  In the evacuated state, no gas is present in the pores of the powder to transfer heat, and therefore only radiative transfer through vacuum between solid particles and conduction through the solid phase contact network are operable.

Other unit-cell micromechanical models not informed by the shape and packing of metal hydride particles, such as that of Zehner, Bauer, and Schl\"{u}nder (see \cite{TsoCEP1987}), yield spurious divergence of $\bar{\kappa}/\kappa_g$ with $\kappa_s/\kappa_g$, as displayed in Fig.~\ref{fig:kks}.  In addition, the ZBS model significantly overestimates conductivity even at moderate $\kappa_s/\kappa_g$.  In contrast, the granular effective medium approximation (GEMA) of Cohen and co-workers \cite{YonJAP1983,SenGeo1981} exhibits consistency with this aspect of transport in metal hydride powder.  GEMA involves the sequential embedding of inclusions within an effective medium formed by a lower density effective medium of inclusions and host material.  A consequence of this sequential application of effective medium theory is the disjointed inclusion topology \cite{YonJAP1983}, a feature that is consistent with the topology of the present jammed systems of faceted particles.  Thus, GEMA provides an excellent estimate of metal hydride conductivity over much of the range of $\kappa_s/\kappa_g$.

\begin{figure}
\centering
\includegraphics[width=\columnwidth]{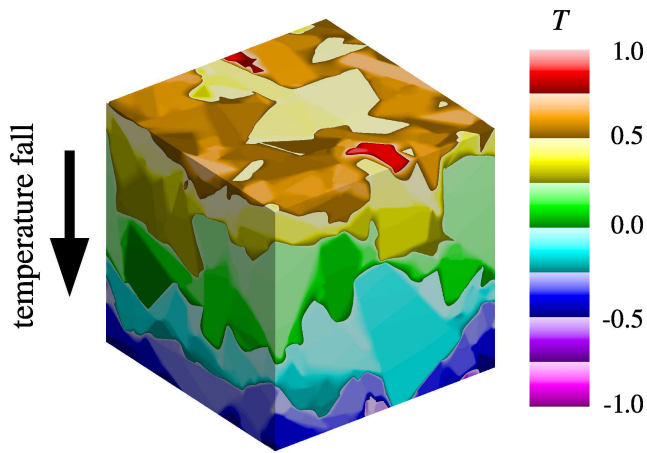}
\caption{Temperature field of the jammed system with $\kappa_s/\kappa_g=10^3$ at $\phi=0.633$.  The field is induced by a unitary periodic temperature fall applied to the primary simulation supercell along the indicated direction.}
\label{fig:tfield}
\end{figure}

Shown in Fig.~\ref{fig:tfield} is the temperature field for the jammed system at a density just below the jamming point.  Solid phase islands are readily apparent in the temperature profile, as they are nearly isothermal.  Because of the lack of connectivity of these islands, $\bar{\kappa}/\kappa_g$ varies little in the high solid conductivity range ($\kappa_s/\kappa_g > 100$).  Transition metals, such as Ni and Ti, are often employed in alloy form as hydrides for H$_2$ storage.  The bulk conductivities among these pure metals span nearly an order of magnitude from 21.9 to 90.7 W/m-K at room temperature \cite{IncFHMT2007}.  The conductivity of pressed solid pellets of H$_2$ storage alloys (e.g., LaNi$_{4.7}$Al$_{0.3}$ and HWT) are of the order of 10 W/m-K (see \cite{HahIJHE1998}).  Bulk H$_2$ gas at 300 K exhibits a conductivity of 0.183 W/m-K \cite{IncFHMT2007}, and consequently $\kappa_s/\kappa_g$ for transition metal hydride powders is expected to be greater than $\sim$100.  With a conductivity contrast ratio this high, enhancement of the intrinsic conductivity of metal hydride particles will yield only marginal enhancement of effective thermal conductivity because of the non-percolating solid-phase granular topology in the powder.

With this information the dependence of $\bar{\kappa}$ on $Kn$ can now be established by expressing $\kappa_g$ as a function of $Kn$.  To do so, we adopt the semi-empirical relation of Kaganer (see \cite{GriIJHMT1999}) employed in the recent metal hydride literature \cite{AsaIJHE2004,UeokaIJHE2007} that is similar to Eq.~\ref{eq:kapg}:

\begin{equation} 
  \kappa_g= \frac{\kappa_0}{1+2bKn}.
\label{eq:kapgKag}
\end{equation} 

\begin{figure}
\centering
\includegraphics[width=\columnwidth]{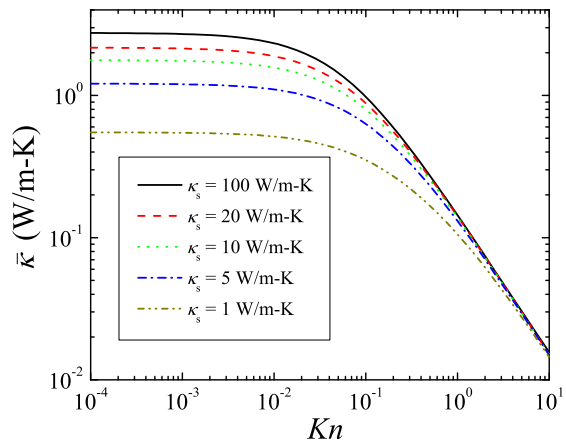}
\caption{Mean effective conductivity $\bar{\kappa}$ for the jammed system at $\phi=0.633$ as a function of Knudsen number $Kn$ for various solid conductivities $\kappa_s$ for H$_2$ gas at 290 K.}
\label{fig:k_vs_Kn}
\end{figure}

\noindent The gas-specific parameter $b$ \nomenclature[b]{$b$}{gas-specific parameter, -} depends on the accommodation coefficient and other parameters \cite{GriIJHMT1999}; here we consider $b$ to have a value of 9.87, as in \cite{AsaIJHE2004,UeokaIJHE2007}.  The bulk conductivity $\kappa_0$ of H$_2$ at 290 K is 0.178 W/m-K, as in \cite{AsaIJHE2004,UeokaIJHE2007}.  As displayed in Fig.~\ref{fig:k_vs_Kn}, moderately to highly conducting solids ($\kappa_s \gtrsim 20$W/m-K) yield very similar effective thermal conductivities.  The Knudsen number $Kn$ scales inversely with gas pressure, and the present results therefore reflect the high-pressure limit of effective thermal conductivity in which the Smoluchowski effect is diminished.

\section{Generalized Structure-Transport Theory}\label{GSTT}

\begin{figure}
\centering
\includegraphics[width=0.5\columnwidth]{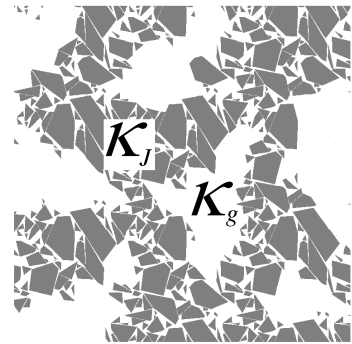} 
\caption{Illustration of a hypothetical cohesive microstructure of metal hydride powder.  Inspired by Refs.~\cite{ValPRL2004,CasPRL2005}, the microstructure is bi-porous with aggregate jammed regions having micro-pores with conductivity $\kappa_J$ \nomenclature[kaJ]{$\kappa_J$}{jammed state conductivity, W/m-K} and void regions forming macro-pores with gas phase conductivity $\kappa_g$.}
\label{fig:cohesion}
\end{figure}

As presented in Sec.~\ref{packing}, the simulated jammed density of particles from the Poisson plane field ($0.665 \pm 0.015$) exceeds that of experimental packings of metal hydride particles (cf., Table~\ref{table:expdensities}).  The primary mechanisms expected to decrease the packing density from the present jamming density are attractive surface interactions between micron-sized metal hydride particles.  Such interactions are known to decrease packing density dramatically \cite{YanPRE2000}, but the non-cohesive jamming point remains a reference state after such systems have undergone compaction \cite{ValPRL2004,CasPRL2005}.  In cohesive packings, agglomerates form as jammed domains with cohesion \cite{ValPRL2004} embedded in void space \cite{CasPRL2005}; the agglomerates form a fractal-like structure with high dimensionality of 2.53 \cite{CasPRL2005}.  Based on this understanding, a low-density packing of metal hydride powder is hypothesized with a granular backbone having local density and structure close to that of the non-cohesive system.  Consequently, the cohesive granular medium is modeled as bi-porous; pores reminiscent of the non-cohesive jammed state are formed with larger pores present due to the steric hindrance of cohesive clusters [Fig.~\ref{fig:cohesion}].

Based on this hypothetical microstructure, the granular effective medium approximation (GEMA) is now employed to model the effective thermal conductivity of cohesive metal hydride powder.  GEMA is based upon a sequential application of Maxwell-Garnett EMA to an effective inclusion-matrix medium (see \cite{TorRHM2002} for a thorough development).  \nomenclature[ka12]{$\kappa_1,\kappa_2$}{phase 1 and 2 conductivities, W/m-K} Firstly, EMA is applied to a pure phase $1$ with perfect continuity and a disjointed spherical inclusion of phase $2$.  To the effective medium phase $e$ formed by these constituents another spherical inclusion of phase $2$ is introduced, and EMA is applied to yield the updated effective medium phase $e$.  This process continues until phase $1$ is fully embedded in phase $2$ and results in a differential equation relating effective conductivity $\kappa_e$ \nomenclature[kae]{$\kappa_e$}{effective medium conductivity, W/m-K} to that of the respective phases and density (see \cite{TorRHM2002}).  The differential equation can be integrated for the process, yielding an implicit expression for the effective conductivity \cite{TorRHM2002}:
 
 
\begin{equation}
  \left(\frac{\kappa_2-\kappa_e}{\kappa_2-\kappa_1}\right)\left(\frac{\kappa_1}{\kappa_e}\right)^{1/3}= 1-\epsilon_1,
\label{eq:GEMA}
\end{equation} 

\noindent where $\epsilon_1$ \nomenclature[ep1]{$\epsilon_1$}{volume fraction of phase $1$, -} is volume fraction of phase $1$.  
The unique feature of this sequential embedding process is the preservation of the continuity of phase $1$, and of the disjointed topology of phase $2$ \cite{YonJAP1983}.  This feature is entirely consistent with our bi-porous hypothetical construction of dense, cohesive metal hydride powder.  Locally jammed aggregates form a continuous phase $1$, while macro-pores form a disjointed phase $2$ in the medium.  In the present cohesive metal hydride powder model, phase $1$ has conductivity $\kappa_1=\kappa_J$, where $\kappa_J$ is the effective conductivity of the non-cohesive jammed system, and phase $2$ has conductivity $\kappa_2=\kappa_g$, where $\kappa_g$ is the gas phase conductivity.  The volume fraction of phase $1$ in the cohesive solid $\epsilon_1$ is expressed in terms of the non-cohesive jammed solid density $\phi_J$ and the cohesive solid density $\phi$ as $\epsilon_1=\phi/\phi_J$.  Substituting these expressions into Eq.~\ref{eq:GEMA}, the GEMA approximation for effective conductivity $\kappa_e$ is obtained:

\begin{equation} 
  \left(\frac{\kappa_g-\kappa_e}{\kappa_J-\kappa_g}\right)\left(\frac{\kappa_g}{\kappa_e}\right)^{1/3}= 1-\phi/\phi_J.
\label{eq:GEMAmh}
\end{equation}

\begin{figure}
\centering
\includegraphics[width=\columnwidth]{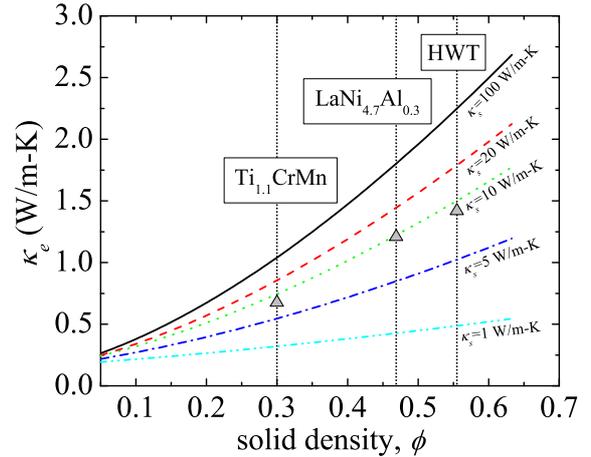}
\caption{Theoretical prediction of effective thermal conductivity for a biporous cohesive metal hydride powder in H$_2$ gas at high pressure (i.e., $Kn \rightarrow 0$).  Triangles represent data for packed metal hydride powder in high pressure H$_2$ gas near 290 K (Ti$_{1.1}$CrMn at 230 bar ($Kn \approx 0.001$) \cite{FlkIJHE2010}; LaNi$_{4.7}$Al$_{0.3}$ at 20 bar ($Kn \approx 0.008$) and HWT at 40 bar ($Kn \approx 0.004$) \cite{HahIJHE1998}).}
\label{fig:biporouskeff}
\end{figure}

Using this effective medium structure-transport theory, the variation of effective thermal conductivity $\kappa_e$ for cohesive metal hydride powder as a function of solid density $\phi$ is predicted in the high gas pressure limit.  To maintain computational efficiency, the properties of the jammed state in the present theoretical model are estimated as that of the system at a density immediately below the jamming threshold at $\phi = 0.633$.  Therefore, the non-cohesive jammed conductivity $\kappa_J$ is estimated by values presented in Fig.~\ref{fig:kks} with $\phi_J=\phi=0.633$.  

The validity of the present theory is supported by the experimental effective conductivity at high gas pressure (small $Kn$) of various hydride powders in Fig.~\ref{fig:biporouskeff}.  A solid-phase conductivity of 10 W/m-K yields good agreement with the present model.  Realizing that $\kappa_s$ for solid pellets of LaNi$_{4.7}$Al$_{0.3}$ and HWT are approximately 10 W/m-K \cite{HahIJHE1998}, the results are especially encouraging.  From the model results in Fig.~\ref{fig:biporouskeff}, the high-pressure conductivity depends weakly on solid conductivity in comparison to the strong dependence on density.  As stated in Sec.~\ref{htsim}, little enhancement of metal hydride powder thermal conductivity can be expected by increasing the intrinsic thermal conductivity of solid metal particles.   Instead, strategies that seek to modify the packing structure of the particles will achieve greater enhancement efficacy.

\begin{figure}
\centering
\includegraphics[width=\columnwidth]{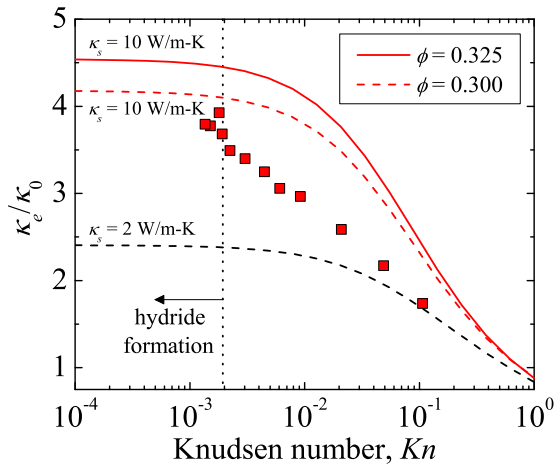}
\caption{Normalized effective conductivity $\kappa_e/\kappa_0$ as a function of Knudsen number $Kn$ of the present theory.  The dashed curves represent values for the density of $\phi = 0.3$ for cycled and packed Ti$_{1.1}$CrMn in Ref.~\cite{FlkIJHE2010}, while the solid curve represents values at higher density ($\phi = 0.325$) as a result of volumetric expansion upon hydriding (see Table~\ref{table:expdensities}).  Curves for $\phi = 0.3$ are shown for various values of $\kappa_s$ as a reference.  Solid squares represent the experimental data of Ref.~\cite{FlkIJHE2010} for cycled Ti$_{1.1}$CrMn, where Knudsen number $Kn$ was estimated with a characteristic pore size equal to the median particle diameter determined in the present work, $D_{50}=3.6 \mu$m.}
\label{fig:flkdata_compare}
\end{figure}

Until this point only the high-pressure limit of gas-phase conductivity in which the Smoluchowski effect is inoperative has been considered.  Employing the approximate results in Fig.~\ref{fig:k_vs_Kn} that incorporate the Smoluchowski effect for the non-cohesive jammed system, the effects of reduced density with the present theory represented by Eq.~\ref{eq:GEMAmh} are incorporated.  In Fig.~\ref{fig:flkdata_compare} these results are compared to experimental data for the high-pressure metal hydride Ti$_{1.1}$CrMn from Ref.~\cite{FlkIJHE2010}.  Theoretical results are displayed for two densities to reflect the effect of volumetric expansion of particles during hydriding.  At low pressure (i.e., high $Kn$) the experimental data follow the theoretical results with a solid conductivity of 2 W/m-K; at high pressure the experimental data follow the theoretical results for solid conductivity of 10 W/m-K, which is closer to our expectations based on the solid conductivity of other intermetallic alloys, e.g., LaNi$_{4.7}$Al$_{0.3}$ and HWT in Ref.~\cite{HahIJHE1998}.  

Therefore, the `solid' phase appears to have a conductivity that depends strongly on H$_2$ gas pressure; physically, this scenario is rather unlikely because deviations with the theoretical results are predominantly within a range of $Kn$ for which hydriding has not initiated for Ti$_{1.1}$CrMn (see dotted line in Fig.~\ref{fig:flkdata_compare}).  Composition changes to the solid phase occur only at very high pressure (low $Kn$).  Instead, the high pressure air titration experiments suggest that the solid is composed of an abundance of internal cracks (cf., Fig.~\ref{fig:ticrmnpsd}).  At intermediate pressures, these cracks reduce the `effective solid' conductivity as a result of low gas density and conductivity within the cracks.  Though this effect is not included herein, the present theoretical model and experimental observations support this conclusion.  

An additional feature reflected by the theoretical model is the spike in conductivity at the initiation of hydriding.  The difference in $\kappa_e$ between the theoretical curves for $\kappa_s=10$~W/m-K at the unhydrided density ($\phi=0.300$) and the approximate hydrided density ($\phi=0.325$) agrees well with the amplitude of the experimental spike.  The elevated conductivity is not maintained after the immediate phase transition, suggesting that dynamic expansion of the powder's free surface occurred after this spike.

\section{Conclusions}

A numerical model for generating metal hydride particles possessing size and shape distributions similar to those observed experimentally has been developed and has been made available for public use on nanohub.org.  Key features of decrepitated metal hydride powders are reflected by the geometric statistical model; this statistical geometric framework may be useful for modeling other strain-induced fragmented shapes (e.g., Si anodes and dried mud).  Subsequent consolidation of sample particle sets generated from this procedure were simulated via energy-based structural optimization, reflecting a structure denser than that of cohesive metal hydride powders.  Effective thermal conductivity of the modeled particle assemblies was simulated, and found to be in excellent agreement with the predictions of granular effective medium theory.  In contrast, the commonly employed Zehner-Bauer-Schl\"{u}nder semi-empirical model deviates strongly from the present results because of the granular nature of metal hydride packings.  Finally, a generalized structure-transport theory is developed based on recent experimental results to model the effective conductivity of metal hydride powders with density below the jamming threshold.   This theory reflects aspects of metal hydride powders of various compositions and packed densities.  These findings suggest that engineering strategies to enhance hydride effective conductivity should seek to increase packing density.  This model for particle shape can also be utilized for simulating metal hydride H$^+$ diffusion kinetics, H$_2$ permeability, and composite effective thermal and mechanical properties.

\section{Acknowledgements}

K.C.S. thanks the U.S. National Science Foundation and the Purdue Graduate School for financial support.  Both authors thank the U.S. National Science Foundation's Office of International Science and Engineering for travel support that enabled illuminating foundational interactions on granular mechanics with Prof. Meheboob Alam of the J. Nehru Centre for Advanced Scientific Research. The authors thank Tyler Voskuilen of the Purdue Hydrogen Systems Laboratory for assistance in the preparation of oxidized hydride samples.  The authors also thank Eric Maw, Claudia Mujat, and Gerald Sando of Malvern Instruments for particle size distribution measurements.  Finally, the authors thank Jayathi Murthy and the PRISM center staff, as well as Phil Cheeseman of the Rosen Center for Advanced Computing, for access to and support of computing resources.





\bibliographystyle{model3-num-names}
\bibliography{../../mainbib.bib}

\end{document}